\documentclass[sigconf]{acmart}

\AtBeginDocument{%
  }

\makeatletter
\renewcommand{\@secfont}{\sffamily\bfseries\section@raggedright\Large}
\makeatother

\renewcommand\footnotetextcopyrightpermission[1]{}
\setcopyright{none}
\copyrightyear{2025}
\acmYear{2025}\copyrightyear{ACM 2025. This is the author's version of the work. It is posted here for your personal use. Not for redistribution. The definitive Version of Record was published in NANOCOM '25: Proceedings of the 12th Annual ACM International Conference on Nanoscale Computing and Communication, http://dx.doi.org/10.1145/{3760544.3764128}}
\acmConference[NanoCom '25]{International Conference on Nanoscale Computing and Communication}{October 23--25, 2025}{Chengdu, China}
\acmBooktitle{International Conference on Nanoscale Computing and Communication (NanoCom '25), October 23--25, 2025, Chengdu, China}
\acmDOI{10.1145/3760544.3764128}
\acmISBN{979-8-4007-2166-3/25/10}

\usepackage{siunitx}
\usepackage{tikz}
\usetikzlibrary{chains, positioning, arrows.meta, decorations.pathreplacing, calc}

\begin{document}

\title{Neural Network based Distance Estimation for Branched Molecular Communication Systems}

\author{Martín Schottlender}
\email{mschottlender@fi.uba.ar}
\affiliation{%
  \institution{Universidad de Buenos Aires, Facultad de Ingeniería}
  \city{Buenos Aires}
  \country{Argentina}
}

\author{Maximilian Schäfer}
\email{max.schaefer@fau.de}
\affiliation{%
  \institution{Friedrich-Alexander Universität Erlangen-Nürnberg}
  \city{Erlangen}
  \country{Germany}
}

\author{Ricardo A. Veiga}
\email{rveiga@fi.uba.ar}
\affiliation{%
  \institution{Universidad de Buenos Aires, Facultad de Ingeniería}
  \city{Buenos Aires}
  \country{Argentina}
}

\begin{abstract}
Molecular Communications (MC) is an emerging research paradigm that utilizes molecules to transmit information, with promising applications in biomedicine such as targeted drug delivery or tumor detection. It is also envisioned as a key enabler of the Internet of BioNanoThings (IoBNT). In this paper, we propose algorithms based on Recurrent Neural Networks (RNN) for the estimation of communication channel parameters in MC systems. We focus on a simple branched topology, simulating the molecule movement with a macroscopic MC simulator. The Deep Learning architectures proposed for distance estimation demonstrate strong performance within these branched environments, highlighting their potential for future MC applications.
\end{abstract}



\maketitle

  \vspace{0.8pt}
  {\small
    \textbf{Author's version}: ©ACM 2025. This is the author's version of the work. It is posted here for your personal use. Not for redistribution. The definitive Version of Record was published in NANOCOM '25: Proceedings of the 12th Annual ACM International Conference on Nanoscale Computing and Communication, http://dx.doi.org/10.1145/{3760544.3764128}
  }

\section{Introduction}

MC is an emerging research paradigm that utilizes molecules as information carriers \cite{farsad16}. This framework applies principles from communication engineering to systems involving molecular transport, prevalent in both biomedical and industrial contexts, with the aim of modeling how information is encoded, transmitted, and decoded through physical and chemical processes. By treating molecular transport as a communication channel, researchers can leverage the intrinsic properties of these systems to develop robust and efficient information transfer mechanisms. This approach supports the development of applications such as \textit{targeted drug delivery} (TDD) or health monitoring \cite{cevallos19}. In the future, MC is expected to enable direct communication among devices within the human body, forming part of the emerging \textit{Internet of BioNanoThings} (IoBNT) \cite{kuscu21,shrivastava21}. IoBNT networks aim to support localized diagnoses and highly personalized treatments at the organ or even cell level.

Among the advantages of IoBNT networks is their capacity to facilitate high-precision biometric analysis, supported by parameter estimation techniques that help characterize communication channels modeled after biological systems. A prominent application enabled by these methods is anomaly detection \cite{etemadi23}. In such scenario, one receiver (Rx), e.g., a wearable device,  simultaneously evaluates signals from different sources, to determine the origin, nature and timing of potential anomalies \cite{pal24}. This enables timely initiation of suitable treatment strategies \cite{ghavami20,mosayebi19}. 

These communication systems must operate reliably in highly branched environments, such as the cardiovascular system with its extensive network of blood vessels \cite{jakumeit24}, or industrial settings involving complex pipeline structures \cite{farsad17}.  Cylindrical geometries are often used in simulations to represent these environments, capturing the structural characteristics of blood vessels and pipelines \cite{wicke18,schaefer20}.

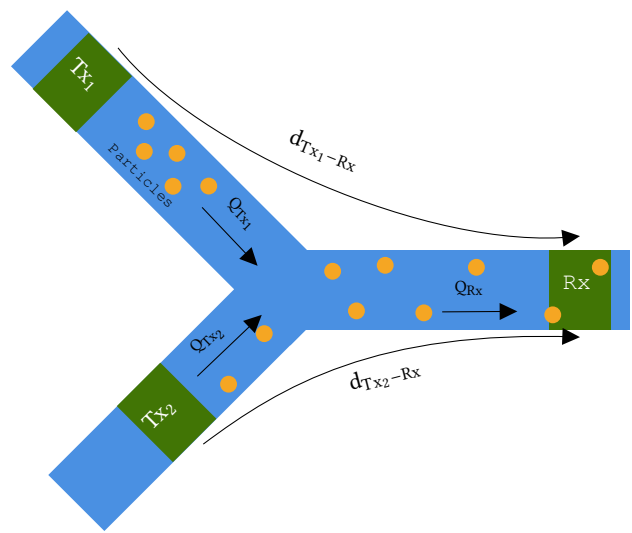
\begin{figure}[htbp]

\begin{tikzpicture}[x=0.75pt,y=0.75pt,yscale=-1,xscale=1]

\draw  [draw opacity=0][fill={rgb, 255:red, 74; green, 144; blue, 226 }  ,fill opacity=1 ] (219.3,-4.99) -- (360.18,135.9) -- (331.9,164.18) -- (191.01,23.3) -- cycle ;
\draw  [draw opacity=0][fill={rgb, 255:red, 74; green, 144; blue, 226 }  ,fill opacity=1 ] (209.83,229.58) -- (315.4,124.01) -- (343.68,152.3) -- (238.12,257.87) -- cycle ;
\draw  [draw opacity=0][fill={rgb, 255:red, 74; green, 144; blue, 226 }  ,fill opacity=1 ] (303.98,116.08) -- (503.22,116.08) -- (503.22,156.08) -- (303.98,156.08) -- cycle ;
\draw  [draw opacity=0][fill={rgb, 255:red, 65; green, 117; blue, 5 }  ,fill opacity=1 ] (462.5,116) -- (493.5,116) -- (493.5,156) -- (462.5,156) -- cycle ;
\draw  [draw opacity=0][fill={rgb, 255:red, 65; green, 117; blue, 5 }  ,fill opacity=1 ] (230.18,6.4) -- (252.1,28.32) -- (223.82,56.6) -- (201.9,34.68) -- cycle ;
\draw  [draw opacity=0][fill={rgb, 255:red, 65; green, 117; blue, 5 }  ,fill opacity=1 ] (244.4,194.82) -- (266.32,172.9) -- (294.6,201.18) -- (272.68,223.1) -- cycle ;
\draw  [draw opacity=0][fill={rgb, 255:red, 245; green, 166; blue, 35 }  ,fill opacity=1 ] (255,51.25) .. controls (255,48.9) and (256.9,47) .. (259.25,47) .. controls (261.6,47) and (263.5,48.9) .. (263.5,51.25) .. controls (263.5,53.6) and (261.6,55.5) .. (259.25,55.5) .. controls (256.9,55.5) and (255,53.6) .. (255,51.25) -- cycle ;
\draw  [draw opacity=0][fill={rgb, 255:red, 245; green, 166; blue, 35 }  ,fill opacity=1 ] (254,66.25) .. controls (254,63.9) and (255.9,62) .. (258.25,62) .. controls (260.6,62) and (262.5,63.9) .. (262.5,66.25) .. controls (262.5,68.6) and (260.6,70.5) .. (258.25,70.5) .. controls (255.9,70.5) and (254,68.6) .. (254,66.25) -- cycle ;
\draw  [draw opacity=0][fill={rgb, 255:red, 245; green, 166; blue, 35 }  ,fill opacity=1 ] (270.5,67.25) .. controls (270.5,64.9) and (272.4,63) .. (274.75,63) .. controls (277.1,63) and (279,64.9) .. (279,67.25) .. controls (279,69.6) and (277.1,71.5) .. (274.75,71.5) .. controls (272.4,71.5) and (270.5,69.6) .. (270.5,67.25) -- cycle ;
\draw  [draw opacity=0][fill={rgb, 255:red, 245; green, 166; blue, 35 }  ,fill opacity=1 ] (268.5,83.85) .. controls (268.5,81.5) and (270.4,79.6) .. (272.75,79.6) .. controls (275.1,79.6) and (277,81.5) .. (277,83.85) .. controls (277,86.2) and (275.1,88.1) .. (272.75,88.1) .. controls (270.4,88.1) and (268.5,86.2) .. (268.5,83.85) -- cycle ;
\draw  [draw opacity=0][fill={rgb, 255:red, 245; green, 166; blue, 35 }  ,fill opacity=1 ] (286.5,83.75) .. controls (286.5,81.4) and (288.4,79.5) .. (290.75,79.5) .. controls (293.1,79.5) and (295,81.4) .. (295,83.75) .. controls (295,86.1) and (293.1,88) .. (290.75,88) .. controls (288.4,88) and (286.5,86.1) .. (286.5,83.75) -- cycle ;
\draw  [draw opacity=0][fill={rgb, 255:red, 245; green, 166; blue, 35 }  ,fill opacity=1 ] (296.5,183.75) .. controls (296.5,181.4) and (298.4,179.5) .. (300.75,179.5) .. controls (303.1,179.5) and (305,181.4) .. (305,183.75) .. controls (305,186.1) and (303.1,188) .. (300.75,188) .. controls (298.4,188) and (296.5,186.1) .. (296.5,183.75) -- cycle ;
\draw  [draw opacity=0][fill={rgb, 255:red, 245; green, 166; blue, 35 }  ,fill opacity=1 ] (314.5,158.25) .. controls (314.5,155.9) and (316.4,154) .. (318.75,154) .. controls (321.1,154) and (323,155.9) .. (323,158.25) .. controls (323,160.6) and (321.1,162.5) .. (318.75,162.5) .. controls (316.4,162.5) and (314.5,160.6) .. (314.5,158.25) -- cycle ;
\draw  [draw opacity=0][fill={rgb, 255:red, 245; green, 166; blue, 35 }  ,fill opacity=1 ] (349,126.75) .. controls (349,124.4) and (350.9,122.5) .. (353.25,122.5) .. controls (355.6,122.5) and (357.5,124.4) .. (357.5,126.75) .. controls (357.5,129.1) and (355.6,131) .. (353.25,131) .. controls (350.9,131) and (349,129.1) .. (349,126.75) -- cycle ;
\draw  [draw opacity=0][fill={rgb, 255:red, 245; green, 166; blue, 35 }  ,fill opacity=1 ] (361,146.75) .. controls (361,144.4) and (362.9,142.5) .. (365.25,142.5) .. controls (367.6,142.5) and (369.5,144.4) .. (369.5,146.75) .. controls (369.5,149.1) and (367.6,151) .. (365.25,151) .. controls (362.9,151) and (361,149.1) .. (361,146.75) -- cycle ;
\draw  [draw opacity=0][fill={rgb, 255:red, 245; green, 166; blue, 35 }  ,fill opacity=1 ] (375.5,123.75) .. controls (375.5,121.4) and (377.4,119.5) .. (379.75,119.5) .. controls (382.1,119.5) and (384,121.4) .. (384,123.75) .. controls (384,126.1) and (382.1,128) .. (379.75,128) .. controls (377.4,128) and (375.5,126.1) .. (375.5,123.75) -- cycle ;
\draw  [draw opacity=0][fill={rgb, 255:red, 245; green, 166; blue, 35 }  ,fill opacity=1 ] (395,147.75) .. controls (395,145.4) and (396.9,143.5) .. (399.25,143.5) .. controls (401.6,143.5) and (403.5,145.4) .. (403.5,147.75) .. controls (403.5,150.1) and (401.6,152) .. (399.25,152) .. controls (396.9,152) and (395,150.1) .. (395,147.75) -- cycle ;
\draw  [draw opacity=0][fill={rgb, 255:red, 245; green, 166; blue, 35 }  ,fill opacity=1 ] (421.5,124.75) .. controls (421.5,122.4) and (423.4,120.5) .. (425.75,120.5) .. controls (428.1,120.5) and (430,122.4) .. (430,124.75) .. controls (430,127.1) and (428.1,129) .. (425.75,129) .. controls (423.4,129) and (421.5,127.1) .. (421.5,124.75) -- cycle ;
\draw  [draw opacity=0][fill={rgb, 255:red, 245; green, 166; blue, 35 }  ,fill opacity=1 ] (460,148.75) .. controls (460,146.4) and (461.9,144.5) .. (464.25,144.5) .. controls (466.6,144.5) and (468.5,146.4) .. (468.5,148.75) .. controls (468.5,151.1) and (466.6,153) .. (464.25,153) .. controls (461.9,153) and (460,151.1) .. (460,148.75) -- cycle ;
\draw  [draw opacity=0][fill={rgb, 255:red, 245; green, 166; blue, 35 }  ,fill opacity=1 ] (484,124.75) .. controls (484,122.4) and (485.9,120.5) .. (488.25,120.5) .. controls (490.6,120.5) and (492.5,122.4) .. (492.5,124.75) .. controls (492.5,127.1) and (490.6,129) .. (488.25,129) .. controls (485.9,129) and (484,127.1) .. (484,124.75) -- cycle ;
\draw    (287.4,94.51) -- (313.33,121.84) ;
\draw [shift={(315.4,124.01)}, rotate = 226.49] [fill={rgb, 255:red, 0; green, 0; blue, 0 }  ][line width=0.08]  [draw opacity=0] (8.93,-4.29) -- (0,0) -- (8.93,4.29) -- cycle    ;
\draw    (408.5,147.25) -- (443.4,147.03) ;
\draw [shift={(446.4,147.01)}, rotate = 179.64] [fill={rgb, 255:red, 0; green, 0; blue, 0 }  ][line width=0.08]  [draw opacity=0] (8.93,-4.29) -- (0,0) -- (8.93,4.29) -- cycle    ;
\draw    (285,179.75) -- (315.33,150.82) ;
\draw [shift={(317.5,148.75)}, rotate = 136.35] [fill={rgb, 255:red, 0; green, 0; blue, 0 }  ][line width=0.08]  [draw opacity=0] (8.93,-4.29) -- (0,0) -- (8.93,4.29) -- cycle    ;
\draw    (245,14) .. controls (283.61,63.25) and (401.61,116.92) .. (475.77,109.25) ;
\draw [shift={(478,109)}, rotate = 173.07] [fill={rgb, 255:red, 0; green, 0; blue, 0 }  ][line width=0.08]  [draw opacity=0] (8.93,-4.29) -- (0,0) -- (8.93,4.29) -- cycle    ;
\draw    (288,214) .. controls (327.8,184.15) and (375.52,156.28) .. (476.47,159.94) ;
\draw [shift={(478,160)}, rotate = 182.25] [fill={rgb, 255:red, 0; green, 0; blue, 0 }  ][line width=0.08]  [draw opacity=0] (8.93,-4.29) -- (0,0) -- (8.93,4.29) -- cycle    ;

\draw (468,128) node [anchor=north west][inner sep=0.75pt]  [color={rgb, 255:red, 255; green, 255; blue, 255 }  ,opacity=1 ] [align=left] {{\fontfamily{pcr}\selectfont Rx}};
\draw (225.11,16.14) node [anchor=north west][inner sep=0.75pt]  [color={rgb, 255:red, 255; green, 255; blue, 255 }  ,opacity=1 ,rotate=-45] [align=left] {$\displaystyle \mathrm{Tx_{1}}$};
\draw (254.56,202.04) node [anchor=north west][inner sep=0.75pt]  [color={rgb, 255:red, 255; green, 255; blue, 255 }  ,opacity=1 ,rotate=-315] [align=left] {$\displaystyle \mathrm{Tx_{2}}$};
\draw (243.27,59.7) node [anchor=north west][inner sep=0.75pt]  [font=\scriptsize,rotate=-44.18] [align=left] {{\fontfamily{pcr}\selectfont Particles}};
\draw (333.24,51.94) node [anchor=north west][inner sep=0.75pt]  [rotate=-21.97] [align=left] {$\displaystyle \mathrm{d_{T}{}_{x_{1} -Rx}}$};
\draw (358.88,177.67) node [anchor=north west][inner sep=0.75pt]  [rotate=-345.56] [align=left] {$\displaystyle \mathrm{d_{T}{}_{x_{2} -Rx}}$};
\draw (304.24,86.41) node [anchor=north west][inner sep=0.75pt]  [font=\scriptsize,rotate=-44.95] [align=left] {$\displaystyle \mathrm{Q_{Tx_{1}}}$};
\draw (413.6,130.07) node [anchor=north west][inner sep=0.75pt]  [font=\scriptsize,rotate=-359.67] [align=left] {$\displaystyle \mathrm{Q_{Rx}}$};
\draw (278.48,164.11) node [anchor=north west][inner sep=0.75pt]  [font=\scriptsize,rotate=-316.18] [align=left] {$\displaystyle \mathrm{Q_{Tx_{2}}}$};

\end{tikzpicture}

\Description{A figure of two horizontal tubes, bending and meeting at a Y-connector, with a main tube leading to a Rx on the right side. The branches have one Tx each, at different distances from the Y-connector. The Txs release particles, which move along the tubes and meet at the main tube, reaching to the Rx.}
\caption{Schematic of the topology of two Txs and one Rx in a branched tube environment. $\mathrm{Q_{Tx_1}}$ and $\mathrm{Q_{Tx_2}}$ are the flow in each branch, and $\mathrm{Q_{Rx}}$ is the flow in the Rx. ${\mathrm{d_{Tx_1-Rx}}}$ is the distance between Tx1 and Rx, and ${\mathrm{d_{Tx_2-Rx}}}$ is the distance between Tx2 and Rx.}
\label{fig:two_tubes_schematic}
\end{figure}

 In many MC systems, distance estimation between the transmitter (Tx) and Rx is a crucial task that enables source localization, anomaly detection, and system calibration. \cite{gulec20}. In branched MC topologies however, this becomes more challenging due to the increased complexity of molecular propagation paths \cite{jakumeit24}. One of these scenarios is depicted in Fig. \ref{fig:two_tubes_schematic}, where an Rx measures the concentration of molecules originating from two Txs, located in two different branches.

 Prior work on parameter estimation in MC has largely focused on classical approaches, including peak-based estimation, maximum likelihood estimation (MLE) \cite{huang22}, least sum of square errors \cite{jamali16} or filters \cite{jing22}. 
 
 Because the analytical modeling of more complex topologies is often infeasible, data-driven techniques have emerged as a possible alternative for the estimation of channel parameters or the design of detection schemes \cite{bartunik22,farsad17}. 
 
 To address these challenges, we propose Recurrent Neural Networks (RNNs) to perform parameter estimation. The algorithms were developed based on the foundational work of Farsad and Goldsmith \cite{farsadgoldsmith}, and were tailored to address the specific challenges of branched structures. 

The remainder of the paper is organized as follows: first, a description of how the simulations are performed for the branched systems, and the analytical model against which the RNN is benchmarked; afterwards, the Neural Network algorithms and the architecture chosen are presented;  later, the results of the most relevant cases are shown; then, a discussion of these results and possible deriving hypotheses; and finally, a brief conclusion about the research.

\section{System model}

In this section, we first present the topology of the model and the overall workflow. Later, we detail the methods and software used to perform the simulations. We also describe the development of the analytical models and how we obtain the resulting parameters for which we will later compare their accuracy with that of the RNN algorithms.

\subsection{Topology}

\begin{figure}[htbp]
\begin{tikzpicture}[>=Stealth, scale=0.8, node distance=1.5cm and 1.8cm, every node/.style={font=\small}]

\node (S0) [draw=none] {$T_{\mathrm{x_1}}$};
\node (S1) [below=0.6cm of S0] {$T_{\mathrm{x_2}}$};
\node (Sn1) [below=1cm of S1] {$T_{\mathrm{x_{N-1}}}$};
\node (Sn) [below=0.6cm of Sn1] {$T_{\mathrm{x_N}}$};

\coordinate (merge) at ($(S0)!0.5!(Sn) + (3,0)$);

\foreach \i in {0,1,n1,n} {
    \coordinate (P\i) at ($(S\i)+(2,0)$);
    \draw[->] ($(S\i)!0.5!(P\i)$) -- ++(0.001,0);
    \draw[->] (S\i) -- (P\i) -- ($(P\i |- merge)$) -- (merge);
}

\node (dots) [below=0.1cm of S1] {$\vdots$};

\node[draw, minimum width=0.8cm, minimum height=0.6cm, right=0cm of merge] (RX) {Rx};

\draw[decorate, decoration={brace, mirror, amplitude=6pt}, thick]
  ([yshift=-0.4cm]Sn.west) -- ([yshift=-2.7cm]RX.east)
  node[midway, below=6pt] {\small Pogona Simulation};

\node[draw, minimum width=1.2cm, minimum height=0.8cm, right=1cm of RX] (SBRNN) {NN};

\draw[->] (RX) -- ++(SBRNN) node[midway, above]{};

\node (O0) [right=1.5cm of SBRNN, yshift=1.2cm] {$\hat{d}_{\mathrm{Tx_1-Rx}}$};
\node (O1) [below=0.3cm of O0] {$\hat{d}_{\mathrm{Tx_2-Rx}}$};
\node (odots) [below=0cm of O1] {$\vdots$};
\node (On1) [below=0.7cm of O1] {$\hat{d}_{\mathrm{Tx_{N-1}-Rx}}$};
\node (On) [below=0.3cm of On1] {$\hat{d}_{\mathrm{Tx_N-Rx}}$};

\draw[->] (SBRNN.east) -- ++(0.8,0) |- (O0.west);
\draw[->] (SBRNN.east) -- ++(0.8,0) |- (O1.west);
\draw[->] (SBRNN.east) -- ++(0.8,0) |- (On1.west);
\draw[->] (SBRNN.east) -- ++(0.8,0) |- (On.west);

\end{tikzpicture}

\Description{A figure of the system model, showing the N sources to the left , joining into the Rx. The results from the Rx are inputs of the SBRNN, which outputs the estimated distance between Tx and Rx for each branch.}
\caption{System model of the proposed distance estimation approach. The molecules which move from the $\mathrm{N}$ sources ($T_{\mathrm{x_1}}$ to $T_{\mathrm{x_n}}$), are received by a single Rx. This is done with the Pogona simulator. The aggregated time-series signal extracted from the Rx is processed by the Neural Network (NN), which outputs the estimated distances between Tx and Rx for each branch.}
\label{system_model}
\end{figure}
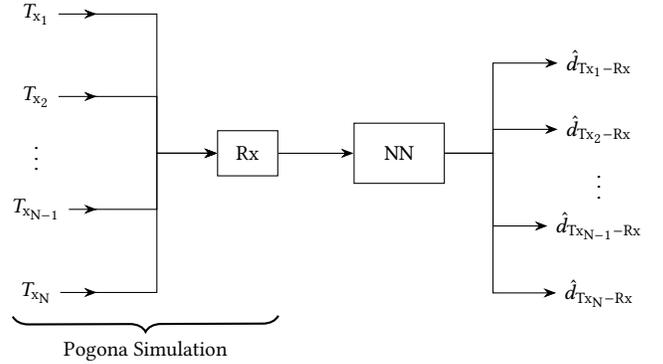

The proposed distance estimation approach is displayed in Fig \ref{system_model}. It involves a system where multiple molecular sources, labeled $T_{\mathrm{x_1}}$ to $T_{\mathrm{x_n}}$, emit molecules that propagate through the branches and are detected by a single Rx. The simulation of molecular propagation and reception is performed using the Pogona simulator, which is described in the next section.

The aggregated signal received at the Rx over time is then processed by a neural network, which is trained to infer the distance between Tx and Rx for each branch $k$, $ d_{\mathrm{Tx,k-Rx}} $.

An analytical mathematical representation has already been developed for a single tube \cite{unterweger18}. However, due to the complex topology present in this scenario, replicating this representation in more intricate branched channels proved challenging. Under these conditions, it was preferred to look for MC simulation approaches.

\subsection{Simulations}\label{sec:simulations}

 The selected modeling environment is \textit{Pogona} \cite{stratmann20}, which is intended as an open source simulator for macroscopic MC, allowing the inclusion of more accurate flow profiles with the use of Computational Fluid Dynamics (CFD) software.

The main concept behind Pogona is the use of a vector field which governs the particles injected by a source into the modeled environment. The construction of this model is done with the Blender\cite{blender18}. These components contain a vector field, calculated by the CFD software OpenFOAM \cite{openfoam07}. The simulations were supported by practical experiments to verify their efficacy for macroscopic MC \cite{bartunik19,unterweger18}. 

We recreated the topology shown in Fig. \ref{fig:two_tubes_schematic} in a Blender model, and then processed the release, movement and sensing of the molecules with Pogona. The particles released at each Tx followed randomized sequences based on an OOK modulation scheme. The simulation parameters are summarized in Table \ref{tab:pogona_parameters}, where $v_{\mathrm{Tx}}$ is the flow speed at the branches where Txs are located, $v_{\mathrm{Rx}}$ is the flow speed at the branch where Rx is located, $\nu$ is the kinematic viscosity of the liquid and $\Delta t_\mathrm{samples}$ is the time between samples. Since for the two-Txs topology, both branches have the same $v_{\mathrm{Tx}}$, and both the two branches and the tube they merge into have the same diameter, the $v_{\mathrm{Rx}}$ is computed as the double of the flow speed at each branch \cite{stratmann20}. The set of simulations included different variations of the distances  between the Txs and the Rx $d_\mathrm{Tx,k-Rx}$, as seen in Table \ref{tab:sets_simulations}. Each of these configurations are simulated with 100 iterations in parallel. Every simulation has a duration of $25 \si{\second}$, with a sequence of 20 symbols of 
 bit at each Tx, where each symbol has a duration of $t_\mathrm{s}=1\ \si{second}$.

\begin{table}[htbp]
\caption{Parameters for the Pogona simulations for 2 Txs and 1 Rx}
\renewcommand{\arraystretch}{1}
\small
    \begin{center}
    \begin{tabular}{clc} 
    \toprule
    Variable & Value\\
    \midrule
        $v_{\mathrm{Tx}}$ & $10$ \si{\frac{ml}{min}} \\
        $v_{\mathrm{Rx}}$ & $20$ \si{\frac{ml}{min}} \\ 
        $\nu$ & $10^{-6}$ \si{\frac{m^2}{s}} \\
        $\Delta t_\mathrm{samples}$ & $0.005\ \si{s}$\\
    \end{tabular}

    \label{tab:pogona_parameters}%
\end{center}
\end{table}

The model assumes the same conditions for both tubes: the flow speed of the liquid, the kinematic viscosity and the angle at which both branches enter the Y-connector to the main tube are all the same. In this case, the estimation models are agnostic towards which particular Tx does the signal come from, meaning that the predicted distances are not strictly assigned to one Tx or the other. Nevertheless, with other topologies, the estimation model could use the differences of each branch to correctly identify which distance belongs to each source. The latter would be a more common in MC since rarely two branches would share exactly the same channel parameters.

Finally, to test the scalability of the estimation models, we tested the performance in a setup with four Txs. This configuration aims to simulate a more challenging environment, where signals from multiple sources can overlap and interact. Due to the increased computational cost of training and evaluating models with four Txs, the number of test cases in this experiment is smaller than in the two-source scenario. Nevertheless, this setup provides valuable initial insights into the model’s behavior under higher source density.

\begin{table}[htbp]
\caption{Simulation configurations used in the experiments with the Pogona simulator, showing the number of cases evaluated for each number of sources.}
\renewcommand{\arraystretch}{1}
\small
    \begin{center}
    \begin{tabular}{clc} 
    \toprule
    Sources&Configuration& \# of cases\\
    \midrule
        1 & $\left\{6\ \mathrm{cm}, 12\ \mathrm{cm},...,\ 24\ \mathrm{cm}\right\}$ & 7  \\
        2 & $\left\{\left[2\ \mathrm{cm},2\ \mathrm{cm} \right], \left[4\ \mathrm{cm},2\ \mathrm{cm} \right] ,..., \left[24\ \mathrm{cm},24\ \mathrm{cm} \right]\right\}$ & 64  \\
        4  & Random combinations & 9  \\
    \end{tabular}

    \label{tab:sets_simulations}%
\end{center}
\end{table}

\subsection{Analytical Models for the Branched Tube Scenario}

The analytical models used to evaluate the performance of the RNNs are based on the stochastic framework introduced in \cite{wicke18}. This model is compatible with the one described in section \ref{sec:simulations}.

The authors in \cite{wicke18} considered the motion of a single particle within a cylindrical structure, in a flow-based regime where laminar flow is dominant. When the molecule spawns on an infinitesimal cross-section of the tube at $x=0$, the probability $P_\mathrm{ob}$ of it reaching a Rx of volume $\mathrm{V_\mathrm{R_{x}}}$ can be obtained as described in \cite{wicke18}[Eq.~(16)]. The parameter $d_{\mathrm{Tx-Rx}}$ denotes the distance between the center of the Tx and the Rx, $v_{\mathrm{eff}}$ is the mean velocity, $L_{\mathrm{Rx}}$ the length across the cross-section of the Rx, $t_{1}$ marks the earliest possible arrival at the Rx, and $t_{2}$ marks the time after which a molecule may have already passed through the entire receiver:

\begin{equation}
    P_{ob}(t) = 
     \begin{cases}
       0 &t \le t_1,\\
       1- \frac{d_{\mathrm{Tx-Rx}} - \frac{L_{\mathrm{Rx}}}{2}}{2 v_{\mathrm{eff}} t} & t_1 < t < t_2,   \\
       \frac{L_{\mathrm{Rx}}}{2 v_{\mathrm{eff}} t} & t \ge t_2 \\
     \end{cases}
\end{equation}

where

\begin{displaymath}
    t_{1,2}=\frac{d_{\mathrm{Tx-Rx}} \mp \frac{L_{\mathrm{Rx}}}{2} }{2 v_{\mathrm{eff}}}
\end{displaymath}

Considering an OOK modulation scheme for the release of particles at Tx, with a series of $\mathrm{K}$ symbols, the number of expected particles at the Rx is as follows:

\begin{equation}
    \mathrm{\overline{N}_{ob}} (t)= N_\mathrm{{Tx}} \sum^{K-1}_{k=0} \mathrm{a}[k] \mathrm{P_{ob}}(t-k T_s)
\end{equation}

\noindent where ${N_\mathrm{{Tx}}}$ is the number of molecules that is released per each symbol, $a[k]$ is the sequence of 1s and 0s, and $T_\mathrm{s}$ is the duration of each symbol.

The case we want to replicate, that in Fig. \ref{fig:two_tubes_schematic}, is a branched system with two branches meeting at a Y-connector. Therefore, to obtain a model for the scenario with two branches, we sum up both contributions:

\begin{displaymath}
 \overline{N}_{ob} (t)= N_{Tx,1} \sum^{K-1}_{k=0} a[k] P_{\mathrm{ob},1}(t-kT_s) +
\end{displaymath}

\begin{equation} \label{eq:poi_estimated_number_2_Tx}
+ N_\mathrm{{Tx,2}} \sum^{K-1}_{k=0} b[k] P_{\mathrm{ob},2}(t-kT_s)
\end{equation}

\noindent where $N_\mathrm{{Tx,1}}$ and $N_\mathrm{{Tx,2}}$ are the number of molecules released at each symbol for the first and second branch, respectively, $P_{\mathrm{ob},1}$ and $P_{\mathrm{ob},2}$ are the probabilities of the molecules from $\mathrm{Tx_1}$ and $\mathrm{Tx_2}$ of reaching the Rx, respectively, and $b[k]$ is the sequence of 1s and 0s of the second Tx. 
The $v_{\mathrm{eff}}$ for each branch is approximated as a weighted average of the flow velocities in the branch and the main tube, taking into account the relative lengths over which the molecule travels.

\begin{figure}[htbp]
\centerline{\includegraphics[width=0.45\textwidth]{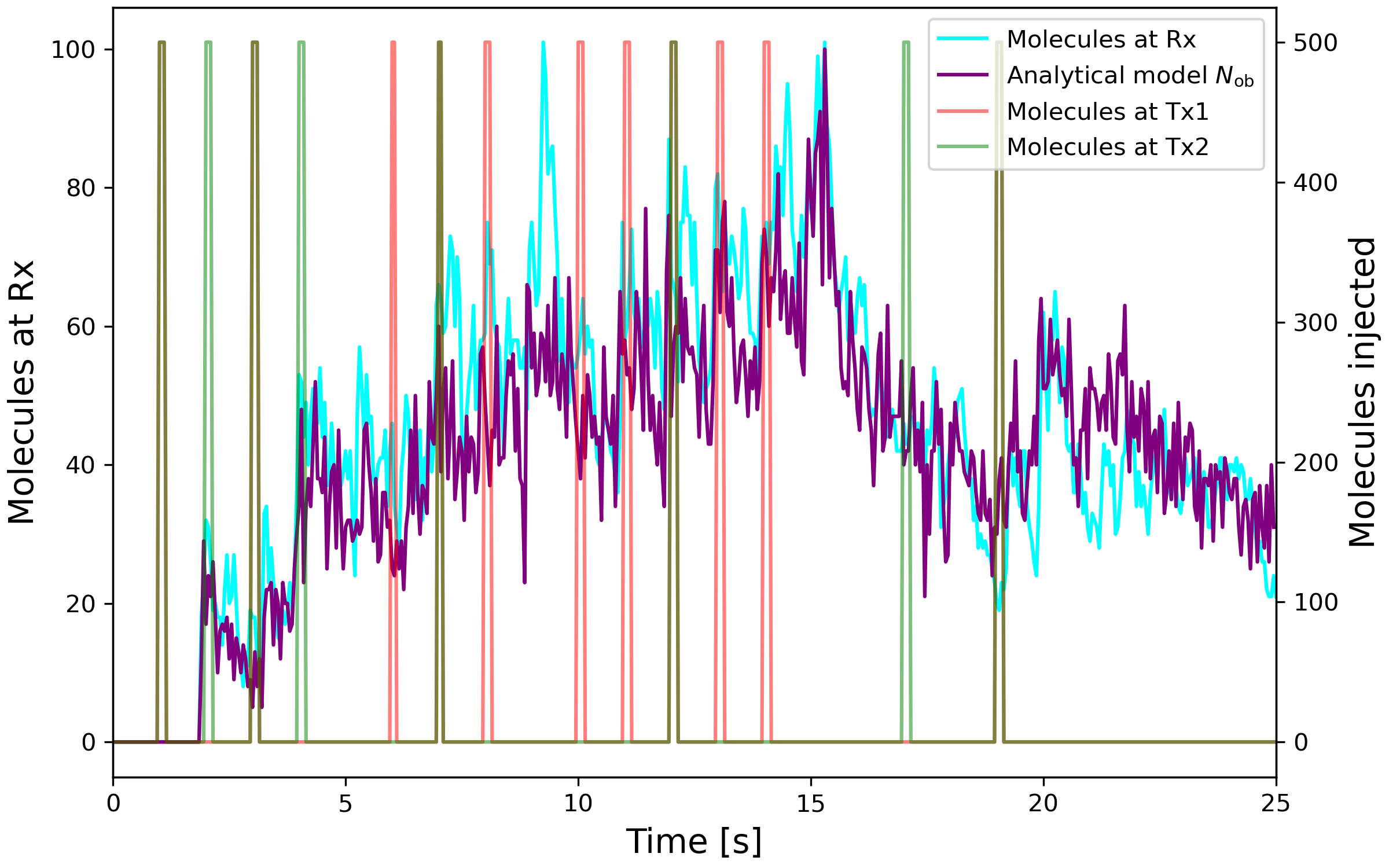}}
\Description{A figure of a tube, with a Tx on the left side showing particles being released along the flow direction, to the right. The flow direction points to the Rx, which is at the right of the tube, at distance d_Tx-Rx from the Tx.}
\caption{Plot of the molecule release at each Tx and the number of molecules at Rx, both for the Pogona simulation and the analytical model, for $ d_{\mathrm{Tx,1-Rx}} = 12\ \si{cm} $ and $ d_{\mathrm{Tx,2-Rx}} = 24\ \si{cm} $ .}
\label{fig:analytical_pogona_molecules}
\end{figure}

A figure to illustrate how the analytical model matches that of Pogona can be seen in Fig. \ref{fig:analytical_pogona_molecules}.

\section{Estimation}

The simulated number of observed molecules in Pogona $\overline{N}_{ob}$ is used as input to the RNNs. During training, the dataset is divided into consecutive batches to expose the model to different temporal starting points and improve generalization. To maximize the memory capacity of the RNN and take advantage of the temporal dependencies, a similar approach as Farsad and Goldsmith \cite{farsadgoldsmith} is utilized: Long Short-Term Memory (LSTM) Bidirectional layers trained with a \textit{Sliding Window} approach. The resulting architecture combines these techniques in the \textbf{Sliding Bidirectional Recurrent Neural Networks} (SBRNN) and is depicted in Fig. \ref{fig:sbrnn_distaince_estimation_architecture}.

\tikzset{every picture/.style={/utils/exec={\sffamily}}}
\usetikzlibrary{shapes.geometric, arrows.meta, positioning, decorations.pathreplacing}

\begin{figure}[htbp]
\begin{tikzpicture}[start chain=going below, scale=0.8, node distance=12pt,
    point/.append style={on chain, join=by {->}, text centered, draw, minimum height=1cm, minimum width=8cm},
    conv/.append style={on chain, join=by {->}, text centered, draw, minimum height=1cm, minimum width=6cm, fill=blue!20},
    activation/.append style={on chain, join=by {->}, text centered, draw, minimum height=1cm, minimum width=6cm, fill=yellow!30},
    pool/.append style={on chain, join=by {->}, text centered, draw, minimum height=1cm, minimum width=4cm, fill=green!30},
    fc/.append style={on chain,  text centered, draw, minimum height=1cm, minimum width=1cm, fill=red!30},
    dropout/.append style={on chain, join=by {->}, text centered, draw, minimum height=1cm, minimum width=5cm, fill=purple!30},
    softmax/.append style={on chain, join=by {->}, text centered, draw, minimum height=1cm, minimum width=4cm, fill=orange!30}
]

    \node[point] {Input Sequence (n=200)};

    \node[conv, text width=6cm, align=center, node font=\normalfont] (conv) {Bidirectional LSTM Layer \\ (Layers=3, n=64x2, tanh activation)};
    \node[activation, text width=6cm, align=center, node font=\normalfont] (dense) {Dense Layer \\ (Layers=5, n=32, ReLu activation)} ;

    \foreach \i/\name in {1/Tx1-Rx, 2/TxN-Rx} {
        \ifnum\i<2
            \node[fc] (fc\i) at ($(dense.south) + (-2.4cm + 1.3*\i cm,-0.3cm)$) {$\hat{d}_{\mathrm{Tx_1-Rx}}$};
            \draw[->] (dense.south) -- (fc\i.north);
        \else\ifnum\i>1
            \node[fc] (fc\i) at ($(dense.south) + (-1.3cm + 1.3*\i cm,-0.3cm)$) {$\hat{d}_{\mathrm{Tx_N-Rx}}$};
            \draw[->] (dense.south) -- (fc\i.north);
        \fi\fi
    }

    \node[text centered] at ($(fc1.east)!0.5!(fc2.west)$) {$\cdots$};

\end{tikzpicture}
\Description{A figure of the layers of the SBRNN with the optimized architecture and hyperparameters. This shows a first layer with the input sequence, which points to the second layer. The latter is a Bidirectional LSTM layer, which in turn points to a Dense Later. This last layer points to each output, two boxes with d_Tx1_Rx and d_TxN_Rx, respectively.}
\caption{Architecture of the SBRNN suited for distance estimation, with the number of layers, the number of neurons $n$ and activation function for each layer, and an output of N Txs.}
\label{fig:sbrnn_distaince_estimation_architecture}
\end{figure}
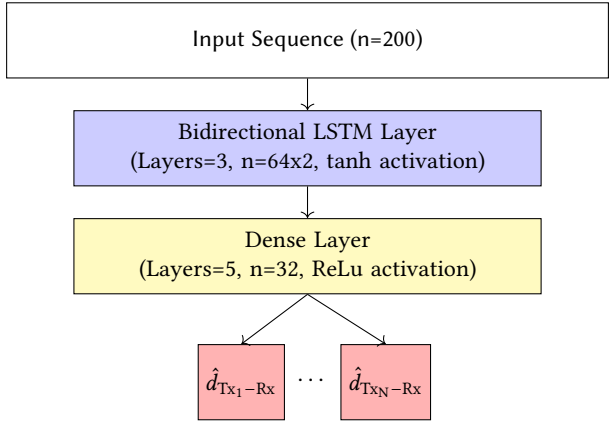

\begin{figure}[htbp]
\centerline{\includegraphics[width=0.4\textwidth]{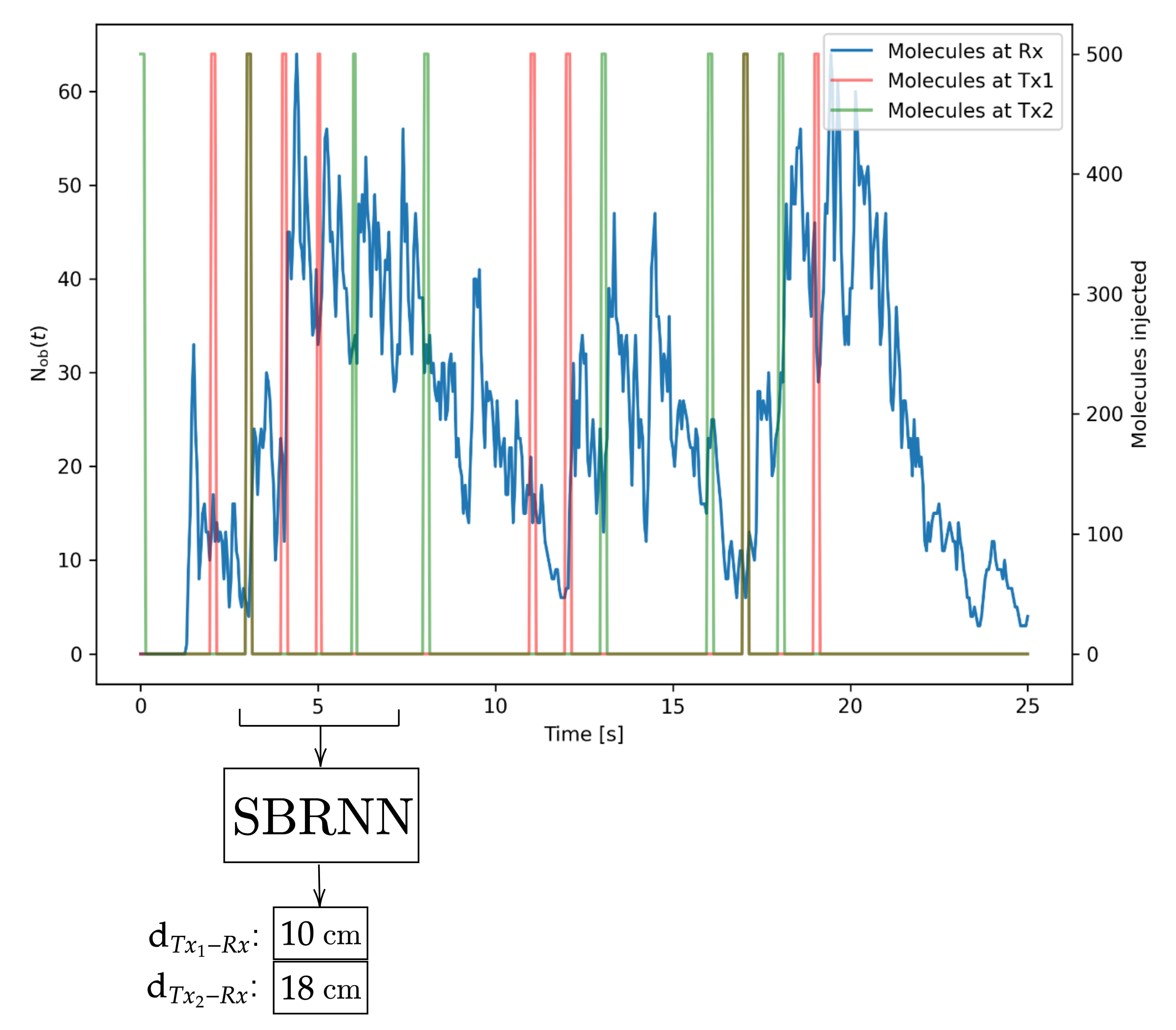}}
\Description{A figure of a graph, showing the number of molecules at the Rx and the pulses of each Tx with respect to time. Under the figure a section of the x-axis, which is time, is bracketed, to represent the sliding window. This bracket points to a square representing the SBRNN. The square points to a table showing the output of the neural network, d_Tx1_Rx and d_Tx2_Rx, which are 10 cm and 18 cm, respectively.}
\caption{Schematic of the neural network algorithm and the sliding window system for the estimation of distance between each Tx and the Rx, for two Txs.}
\label{fig:sbrnn_distance_schematic}
\end{figure}

The algorithms developed in \cite{farsadgoldsmith} were designed for sequence detection for a single source of particles in an unbounded 3-D environment. Since in our case the SBRNN architecture is used for parameter estimation in branched systems, it is adapted to handle a regression problem by adding dense layers with ReLu activation. The output of the SBRNN is then the estimated $d_{\mathrm{Tx,k-Rx}} $, as seen in Fig. \ref{fig:sbrnn_distance_schematic}. Additionally, the schematic shows the window used for the input to the SBRNN, overlapped with the sequences of pulses generated by the two Txs. This window spans $\mathrm{N}=8$ bits and includes 2 additional padding bits to account for potential delays in particle arrival at the Rx when the $d_\mathrm{Tx-Rx}$ is large. To reduce model complexity, the input signal is then undersampled by a factor of 10, resulting in a final input size of: $$\frac{\mathrm{N_{bits}}\ \mathrm{BR}+\mathrm{Padding}}{10.\Delta t} = 200\ \mathrm{samples}$$

The parameters for the SBRNN in Fig. \ref{fig:sbrnn_distaince_estimation_architecture} were selected with hyperparameter tuning, using the RayTune library to aid in finding the best number of layers and neurons. The network was trained using the Mean Squared Error (MSE) as the loss function and the Adam optimizer.

For the analytical baseline, a Maximum Likelihood approach was applied to estimate each $d_{\mathrm{Tx,k-Rx}}$, using \eqref{eq:poi_estimated_number_2_Tx} and the results of the Pogona simulations at each timestep.

For the scenario with four sources, the final layer was configured to output four separate distance estimates, one for each Tx.

\section{Numerical evaluation}

In this section, we present the results of the performance evaluation for two cases: a) the model with two Txs, as in Fig. \ref{fig:two_tubes_schematic}, and b) the model with four Txs. We also explain the metrics used to test the models and provide possible hypotheses deriving from the results. 

The datasets are divided into training, validation and test sets using a 70-20-10 \% split, respectively. For the test sets, the predicted distances $d_{\mathrm{Tx,k-Rx}}$ are compared to the ground truth distances between Tx and Rx. The mean predicted distance $\bar{d}_{\mathrm{Tx,k-Rx}}$ and standard deviations are obtained to measure the accuracy and precision of the estimation for each particular configuration. The performance of the algorithm for each configuration is also assessed by the percentage of test cases in which the difference between the predictions and the ground truth (GT) is smaller than a certain Relative Error (RE):
$$\mathrm{RE=\left|\frac{Predicted\ Values-GT}{GT}\right|}100\%$$

\subsection{Two Source Model}

In this section, we consider the scenario shown in Fig. \ref{fig:two_tubes_schematic}, consisting of two branches with two Txs and one Rx. Table~\ref{tab:two_tubes_SBRNN_analytical} shows the comparative performance of the SBRNN and the analytical model in the two-tube configuration. 

\begin{table}[htbp]
\caption{Relative error for all sets of $d_{\mathrm{Tx-Rx}}$ of a single tube model for the SBRNN and the analytical model}
\renewcommand{\arraystretch}{1}
\small
    \begin{center}
    \begin{tabular}{clc} 
    \toprule
    Metric & SBRNN & Analytical Model\\
    \midrule
        \textbf{\textit{RE \textless 5 \%}}  & 67.70  \% & 74.55  \%  \\ 
        \textbf{\textit{RE \textless 10 \%}}  & 79.84 \%  & 79.25  \%  \\ 
        \textbf{\textit{RE \textless 20 \%}}  & 89.52 \%  & 86.40  \%    \\
    \end{tabular}

    \label{tab:two_tubes_SBRNN_analytical}%
\end{center}
\end{table}

These results indicate that, while the analytical model is slightly better at precise estimation in this setting, the SBRNN is more robust overall, especially at moderate to high error tolerances. This suggests the SBRNN may generalize better in scenarios where the input signals are noisier or where asymmetries between branches make analytical models less accurate.

\begin{figure}[htbp]

\centerline{\includegraphics[width=0.35\textwidth]{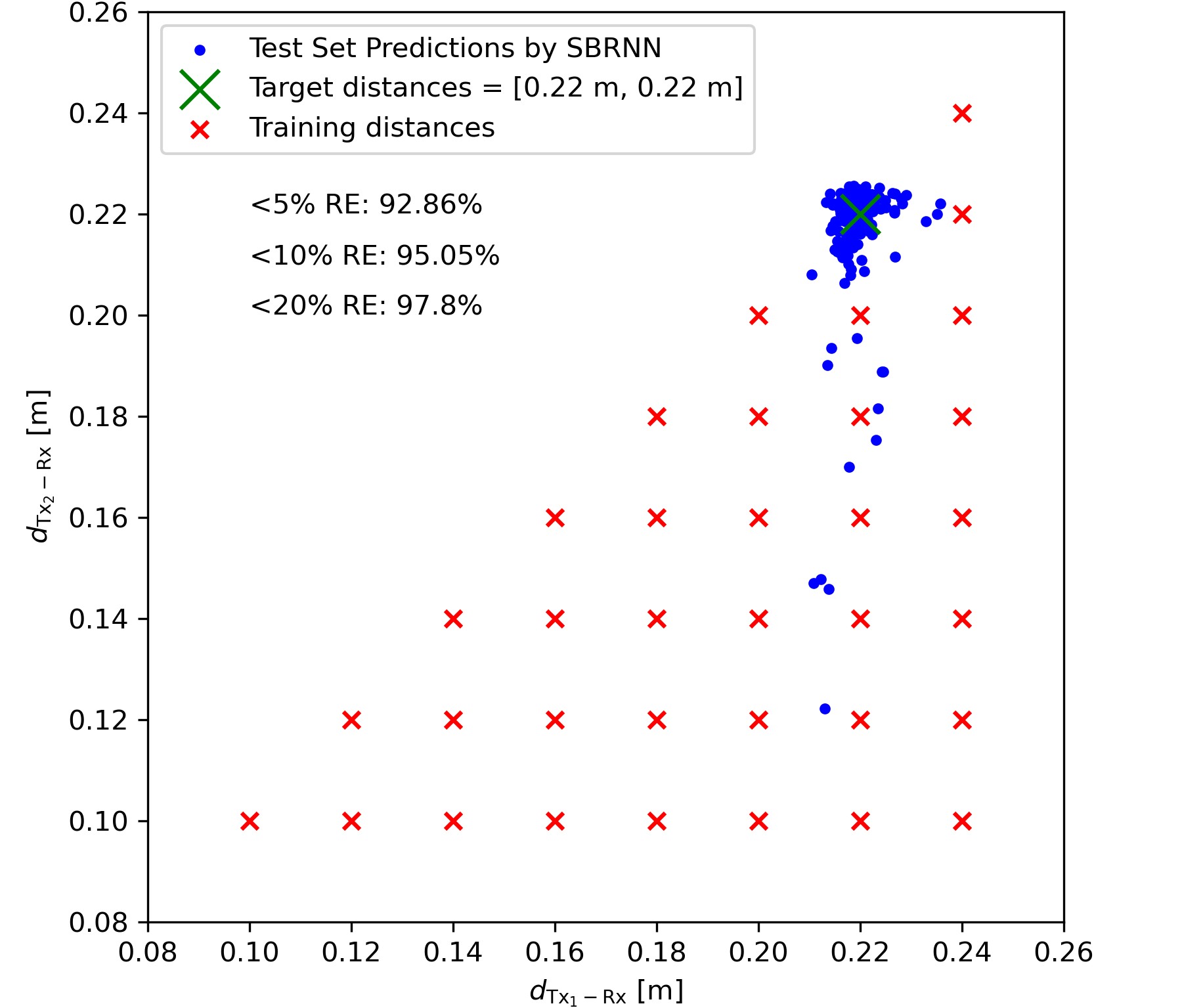}}
\Description{A graph of d_Tx_1 against d_Tx_2 showing all the ground truth dataset distance pairs in red, except the one that is tested, at [22 cm, 22 cm], which is the target and is in green. There are dots around the target, which represent all the estimations done by the NN. 85.64 \% of these are within 5 \% of RE, 96.41 \% within 10 \% and 99.49 \% within 20 \%.}
\caption{Scatter plots of the test results for the SBRNN distance estimation algorithm for $d_{\mathrm{Tx1-Rx}} = 22\ \si{cm},  \ d_{\mathrm{Tx2-Rx}} = 22\ \si{cm} $.}
\label{fig:d_tx1_tx2_graph_1}
\end{figure}

\begin{figure}[htbp]
\centerline{\includegraphics[width=0.35\textwidth]{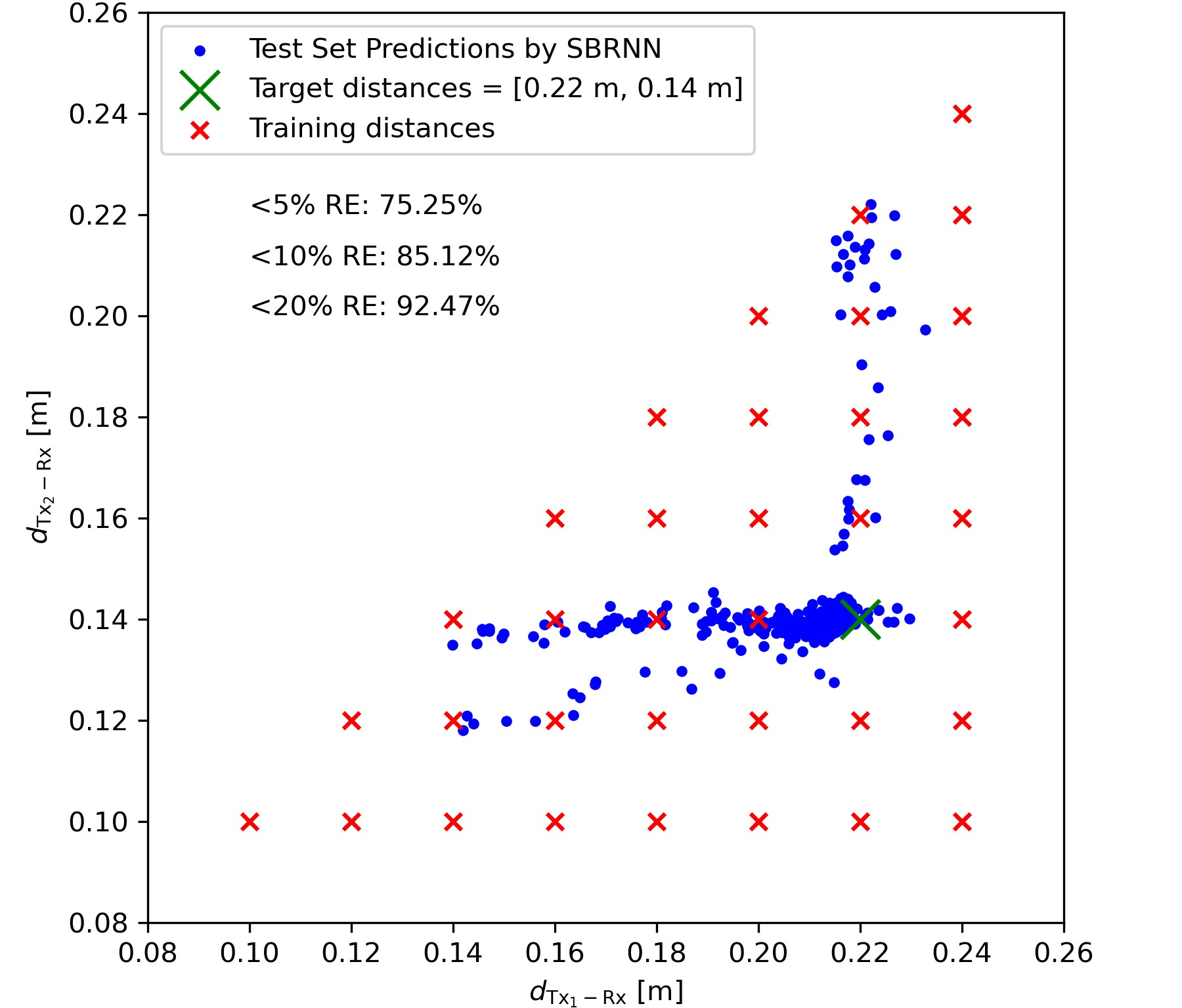}}
\Description{A graph of d_Tx_1 against d_Tx_2 showing all the ground truth dataset distance pairs in red, except the one that is tested, at [22 cm, 14 cm], which is the target and is in green. There are dots around the target, which represent all the estimations done by the NN. 75.43 \% of these are within 5 \% of RE, 83.55 \% within 10 \% and 90.6 \% within 20 \%.}
\caption{Scatter plots of the test results for the SBRNN distance estimation algorithm for $d_{\mathrm{Tx1-Rx}} = 22\ \si{cm},  \ d_{\mathrm{Tx2-Rx}} = 14\ \si{cm} $.}
\label{fig:d_tx1_tx2_graph_2}
\end{figure}

 Examples are shown in Figs. \ref{fig:d_tx1_tx2_graph_1} and \ref{fig:d_tx1_tx2_graph_2}, which illustrate how the test results for two selected sets deviate from the ground truth, The ground truth distances $[d_{\mathrm{Tx,1-Rx}},d_{\mathrm{Tx,2-Rx}}]$ from the training set are also included for visual reference.

\begin{figure}[htbp]
\centerline{\includegraphics[width=0.35\textwidth]{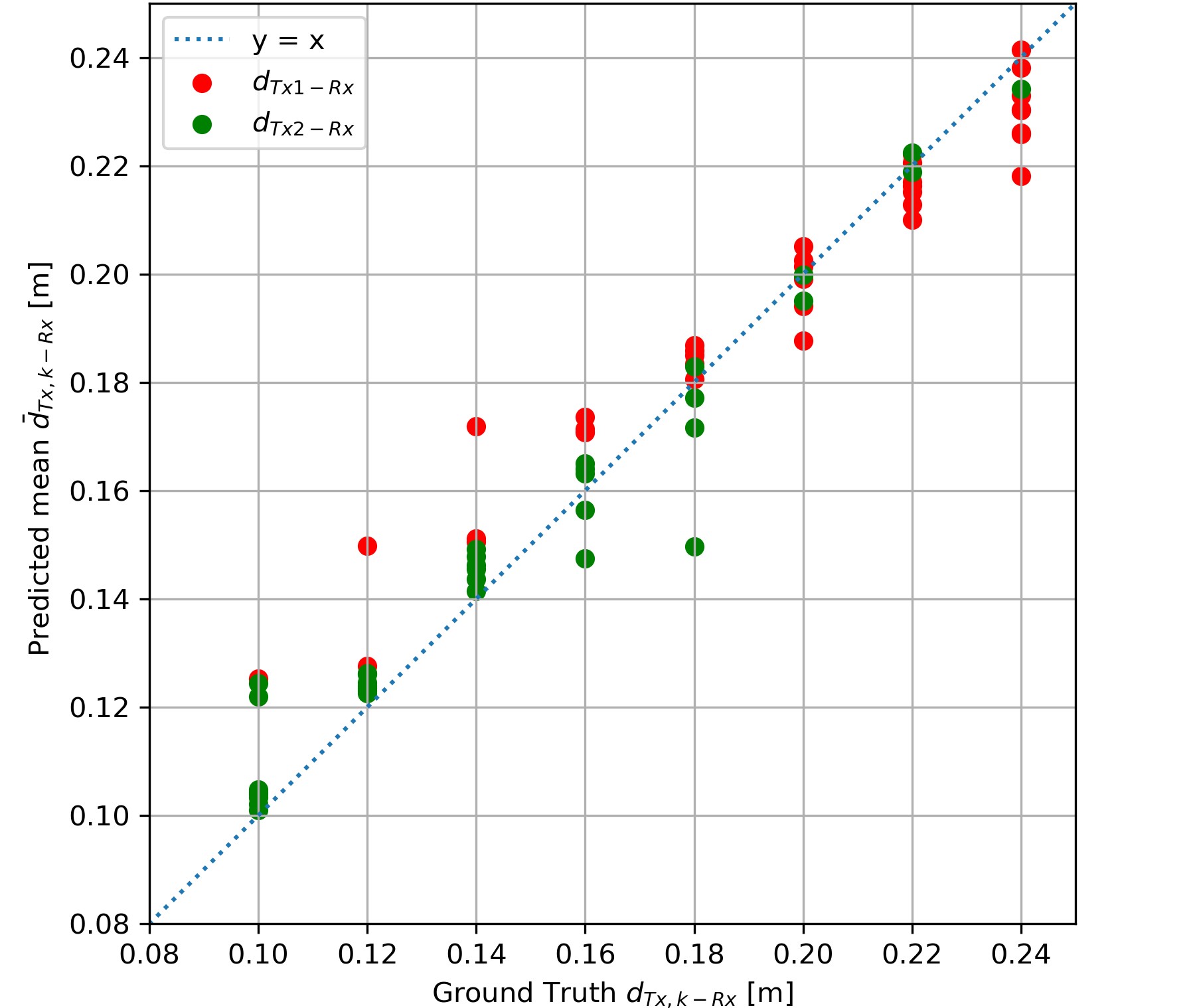}}
\Description{A graph of the ground truth d_Tx_k-Rx against the mean predicted d_Tx_k-Rx showing for each test set the performance with red dots for branch 1 and green dots for branch 2. There is also a y=x line to see how far of the correct result each test set fared. These show that the smaller the d_Tx_k-Rx, the difference is more noteworthy, but the larger, the more they get closer to y=x.}
\caption{Predicted mean $\bar{d}_{\mathrm{Tx,k-Rx}}$ for each configuration tested against the Ground Truth for the two sources case.}
\label{fig:d_p_vs_gt_d_Tx1}
\end{figure}

 Figure~\ref{fig:d_p_vs_gt_d_Tx1} presents the mean predicted distances $\bar{d}_{\mathrm{Tx,k-Rx}}$ against the ground truth distances $d_{\mathrm{Tx,k-Rx}}$ for both Txs. Each red dot corresponds to predictions for Tx1, while green dots represent Tx2, respectively. The diagonal dotted line ($y = x$) denotes a perfect match between the mean prediction and the ground truth, serving as visual aid.

For the case when both Txs have the same distance from the Rx (e.g., in Fig. \ref{fig:d_tx1_tx2_graph_1}), the predicted distances for both branches align well with the ground truth, showing minimal deviation. This is particularly visible in the central cluster of points, where predictions for both Tx1 and Tx2 almost coincide with the dotted line.

However, when the distances of the two Txs to the Rx deviate stronger, the performance worsens. Specifically, the predictions for the Tx further from the Rx tend to underestimate the true distance, while the closer Tx's predictions remain more accurate. This is evident in the outer regions of the plot, where green and red dots begin to diverge from the dotted line asymmetrically.

A potential reason for this could be overshadowing: the stronger signal from the Tx closer to the Rx dominates the received signal, making it difficult for the SBRNN to accurately separate the contributions from each Tx. As a result, the model struggles to maintain precision when one signal is significantly weaker.

This demonstrates that the SBRNN performs well when both Txs are approximately equidistant from the Rx, but performance degrades when there is a significant distance imbalance. This insight could guide training procedures to compensate for this source of error.

\subsection{Multiple Sources}

In this section, we consider the scenario with four branches with four Txs and one Rx. The RE of the SBRNN's estimations across all four Txs and test cases is summarized in Table \ref{tab:four_tubes_SBRNN}.

\begin{table}[htbp]
\caption{Relative error for all sets of $\mathbf{d}_{\mathbf{Tx_{k}-Rx}}$ of a four-Tx model for the SBRNN.}
\renewcommand{\arraystretch}{1}
\small
    \begin{center}
    \begin{tabular}{clc} 
    \toprule
    Metric & Results\\
    \midrule
        \textbf{\textit{RE \textless 5 \%}}  & 63.81  \%    \\
        \textbf{\textit{RE \textless 10 \%}}  & 66.41 \%  \\ 
        \textbf{\textit{RE \textless 20 \%}}  & 78.95 \%   
    \end{tabular}
\label{tab:four_tubes_SBRNN}%
\end{center}
\end{table}
 The $d_{\mathrm{Tx,k-Rx}}$ were randomly selected for each branch, generating variability across the experiments and testing the network’s ability to generalize across a broader range of spatial configurations.

From the results in this table, we observe that while these figures are lower than those achieved in the one- and two-source scenarios, they still indicate that the proposed approach retains a reasonable degree of accuracy even with increasing number of Txs.

This performance drop could be attributed to the growing complexity of the signal space: with four sources, the overlapping and interference between received signals becomes more pronounced. In some instances, the network is able to estimate one or more distances accurately while missing others, suggesting that the model partially resolves the problem, but struggles to consistently decode all Txs in denser configurations.

Despite these challenges, the SBRNN still provides acceptable estimates in the majority of cases. The fact that nearly 80\% of the predictions fall within a 20\% relative error range demonstrates the potential of the model to handle multi-source topologies.

\section{Conclusion and future directions}

In this paper, we proposed a novel model using an adapted SBRNN to estimate the distance between Txs and one Rx in branched systems using MC. When comparing the two-source model with the benchmark, the results suggest that this data-driven approach using the adapted SBRNN is a promising method for measuring distances between Txs and Rx in Molecular Communication systems. This approach is especially remarkable considering that the neural network, unlike a more classical analytical approach, does not receive previous parameters (such as the number of particles released from each Tx, or the flow speed of each branch). These findings are further supported with simulations with multiple Txs in branched environments.

The adapted SBRNN achieves a performance that encourages further work. It provides a novel architecture that accounts for multiple sources in a branched topology, and the new parameter estimation NN offers an innovative way to obtain channel characteristics. Future research will aim to improve the estimation of other channel parameters, analyze other topologies, and include the influence of different physical effects. The influence of diffusion in particular could be investigated, since the Pogona simulator does not include it. Also, more complex analytical models and experimental tests on measured data are planned for future studies.

\begin{acks}
The work of Maximilian Schäfer is partly funded by the Deutsche Forschungsgemeinschaft (DFG, German Research Foundation) – GRK 2950 – Project-ID 509922606, and partly by the European Union’s Horizon Europe – HORIZON-EIC-2024-PATHFINDEROPEN-01 under grant agreement Project N. 101185661. The work of Ricardo A. Veiga is partly supported by UBACYT 20020220200075BA (Universidad de Buenos Aires). 

Gratitude is extended to BAYLAT, who provided the financial support necessary to carry out the research for this project. The use of the Tupac CSC-Conicet cluster for performing calculations is also greatly appreciated.
\end{acks}

\bibliographystyle{ACM-Reference-Format}
\bibliography{bibfile}

\end{document}